# A mempolar transistor made from tellurium

*Yifei Yang†, Lujie Xu†, Mingkun Xu†, Huan Liu, Dameng Liu, Wenrui Duan, Jing Pei, Huanglong Li\**

Y. Yang, M. Xu, J. Pei, H. Li
Department of Precision Instrument, Center for Brain Inspired Computing Research, Tsinghua University, Beijing, 100084, China.
E-mail: li_huanglong@mail.tsinghua.edu.cn

L. Xu, W. Duan
School of Instrument Science and Opto Electronics Engineering, Beijing Information Science and Technology University, Beijing, 100192, China.

L. Xu
Application Technology Department, Dongfang Jingyuan Electron Limited, Beijing, 100176, China.

H. Liu, D. Liu
State Key Laboratory of Tribology, Tsinghua University, Beijing, 100084, China.

H. Li
Chinese Institute for Brain Research; Beijing, 102206, China.

Keywords: transistors, memristors, tellurium, reconfigurable polarity, mempolar, ternary content-addressable memories, regularization methods

**Abstract**
The classic three-terminal electronic transistors and the emerging two-terminal ion-based memristors are complementary to each other in various nonconventional information processing systems in a heterogeneous integration approach, such as DRAM/storage-class memory hierarchy, hybrid CMOS/memristive neuromorphic crossbar arrays, and so on. Recent attempts to introduce transitive functions into memristors have given rise to gate-tunable memristive functions, hetero-plasticity and mixed-plasticity functions. However, it remains elusive under what application scenarios and in what ways transistors can benefit from the incorporation of ion-based memristive effects. Here, we introduce a new type of transistor named 'mempolar transistor' to the transistor family that has included the well-known unipolar and ambipolar transistors. As its name suggests, mempolar transistor has polarity with memory, reminiscent of memristor having resistance with memory. Specifically, its polarity can be converted reversibly, in a nonvolatile fashion, between n-type and p-type depending on the history of the applied electrical stimulus. This is achieved by the use of the emerging semiconducting tellurium as the electrochemically active source/drain contact material, in combination with monolayer two-dimensional $MoS_2$ channel, which results in a gated lateral $Te/MoS_2/Te$ memristor, or from a different perspective, a transistor whose channel can be converted reversibly between n-type $MoS_2$ and p-type Te. With this unique mempolar function, our transistor holds the promise for reconfigurable logic circuits and secure circuits, addressing a fundamental limitation in previous implementations, that is, polarity reconfiguration was volatile and required additional gate terminals. In addition to this manifest advantage, we propose and demonstrate experimentally, a ternary content-addressable memory made of only two mempolar transistors, which used to require a dozen normal transistors, and by simulations, a device-inspired and hardware matched regularization method 'FlipWeight' for training artificial neural networks, which can achieve comparable performance to that achieved by the prevalent 'Dropout' and 'DropConnect' methods. This work represents a major advance in diversifying the functionality of transistors.

## 1. Introduction

Polarity is one of the most fundamental aspects of a transistor's identity, according to which transistors are most famously categorized into either n-type or p-type. The pairing of n-type and p-type transistors has resulted in complementary metal-oxide-semiconductor (CMOS) technology, the leading technology for digital integrated circuits over the past four decades. The polarity of a transistor is a reflection of the comprehensive effect of its entire materials system, including the channel semiconductor, impurities, source/drain contact, and so on. Typically, polarity identity is determined at the fabrication stage and cannot be altered afterwards.

Traditionally, the performance improvements of the integrated circuits have depended almost solely on the miniaturization of transistors (Moore's Law). As miniaturization is approaching its physical limits, transistor scaling is delivering performance improvements at a slower pace. However, software is evolving extremely rapidly with emerging applications[1,2], such as artificial intelligence. Given this, the hardware that cannot adapt to software will suffer from a short lifecycle and high nonrecurring engineering cost[3]. In addition, with the ever increasing density of transistors in microprocessors, power dissipation has become a huge factor challenging the continuous success of the CMOS technology. Overall, both flexibility and energy efficiency have become main criteria for computing fabrics. In this context, incorporation into devices of functionalities that do not necessarily scale according to Moore's Law but provide additional value in different ways has become increasingly pursued by semiconductor industry and academia[4].

Different from transistor in many key aspects, memristor, experimentally discovered less than two decades ago[5], is one such emerging device that holds great promise for low-power and adaptive electronics. Memristor is best-known as a two-terminal resistor with long-term memory. In contrast to the transient electronic switching (volatile) during the operation of transistor, enduring atomic structure change in the switching medium can be elicited when suitable electrical stimulus is applied to memristor, giving rise to nonvolatile reconfigurability of the resistance state. This unique property of memristor has made it a key complementary device to transistor in forming DRAM/storage-class memory hierarchy in conventional von Neumann architecture and in enabling the emerging computing paradigms, such as in-memory computing and neuromorphic computing[6-8], where transistors and memristors are integrated heterogeneously.

As one step further, attempts have been made to introduce transitive functions into memristors. In such devices, additional control gates are positioned aside and electrically insulated from the memristive channels[9]. These have given rise to gate-tunable memristive functions[10-12], enabling the emulation of hetero-plasticity[13-15].

Apparently, memristors and memristor-centered circuits benefit from the additional transitive functions to further enhance their flexibility and augment their functionalities[16,17].

At this point, an interesting question arises as to under what application scenarios and in what ways transistors can benefit from the incorporation of ion-based memristive effects. Here, we propose and demonstrate that the fusion of transistive and memristive functions can result in a novel type of transistor which we name 'mempolar transistor'. As its name suggests, mempolar transistor has polarity with memory, reminiscent of memristor. Specifically, its polarity can be converted reversibly, in a nonvolatile fashion, between n-type and p-type depending on the history of the applied electrical stimulus. Mempolar transistors bring major improvements over the existing reconfigurable transistors[18-21] in the following key aspects: first, mempolar transistors are more energy-efficient because polarity conversion takes place in a nonvolatile fashion, whereas in previous devices control-gate voltages have to be applied persistently; second, mempolar transistors have smaller footprints and less fabrication complexity because they retain the classic three-terminal structures without the addition of more control gates.

To enable mempolarity, a key design idea is to first make a two-terminal memristor whose two resistance states are of n-type and p-type semiconductivity, respectively, then introduce a third terminal to provide gate control over the memristive channel in either polarity, as in a normal transistor. To this end, we choose monolayer two-dimensional (2D) $MoS_2$ as the pristine n-type memristive medium with lateral p-type tellurium (Te) electrodes, making a two-terminal lateral $Te/MoS_2/Te$ device. The memristive function of this device can be foreseen (demonstrations provided) based on the previous demonstrations of the electrochemically activity of Te and other memristors made from it[22-25]. With a back gate, a mempolar transistor that can be reconfigured between an n-type $MoS_2$ transistor and a p-type Te transistor is created. In addition to the aforementioned apparent advantages over the existing reconfigurable transistors in logic and secure circuits, our mempolar transistor also demonstrate additional value in other applications, such as ternary content-addressable memory (TCAM) cell made of two mempolar transistors, which used to require a dozen normal transistors. Inspired by the device properties, a hardware-matched regularization method for mitigating the over-fitting problems in artificial neural networks is also developed, which can achieve comparable performance to that achieved by the prevalent 'Dropout' and 'DropConnect' methods. Our proposed mempolar transistor is a valuable addition to the transistor family, enabling nonvolatile fine-grain reconfigurability and supporting general-purpose hardware design.

## 2. Results and discussion
### 2.1. Mempolar function and its mechanism
The schematic structure and the optical image of the mempolar transistor are shown in figure 1a and 1b, respectively. To fabricate the device, a monolayer $MoS_2$ flake (the monolayer characteristics verified by Raman spectroscopy are shown in supplementary figure S1) is mechanically exfoliated onto a heavily p-type doped Si substrate with 300-nm-thick thermally oxidized $SiO_2$. The two Te electrodes are then deposited by magnetron sputtering as the source and drain terminals, followed by the deposition of platinum protective layers (see Methods).

The n-type transfer characteristic of the as-fabricated device is depicted in figure 1c. Under a constant drain-source voltage ($V_{ds}=V_d-V_s$) of +1 V, the drain current ($I_d$) is increased by about 200 times as $V_g$ is swept from -10 V to +10 V. Certain degree of clockwise hysteresis can also be observed as $V_g$ is swept back and forth, indicating that trapping/detrapping of electrons in/from the gate oxide ($SiO_2$) takes place during the course of $V_g$ sweeping. To elicit polarity conversion, a sufficiently large $V_{ds}$ pulse of +30 V in amplitude and 30 s in duration is applied to the device (under $V_g=0$ V). As aforementioned, the Te electrode under negative bias (source electrode) with respect to the other can be electrochemically reduced. The induced $Te^{2-}$ anions may migrate towards the counter electrode where they will be re-oxidized to elemental Te. As this electrochemical process proceeds, this accumulated elemental Te may eventually bridge the source and drain electrodes by forming local Te filament or wider Te sheet. Because Te is a narrow-bandgap semiconductor with native p-type conductivity, we have previously exploited this electrochemical mechanism to enable filamentary resistance switching in vertical memristors where wider-bandgap (more insulating) dielectrics are sandwiched between the Te electrodes[23]. Here, the formation of Te sheet of the width as close to that of the channel as possible is preferred in order to best maintain the channel geometry. Figure 1d shows the transfer curve of the device after large $V_{ds}$ is applied. It is seen that $I_d$ is now decreased with increasing $V_g$, verifying that transistor polarity is conversed to p-type. With the conversion of polarity, the hysteresis in the transfer curve also changes direction from clockwise to anti-clockwise. The p-type transfer characteristics can still be reproduced after one month without degradation, confirming the nonvolatile nature of the polarity conversion. The p polarity can be reversibly switched back to n polarity by applying a $V_{ds}$ pulse of -30 V under $V_g=+5$ V. Gate biasing here reduces the number of free hole carriers in the p-type Te channel and thus mitigates channel field screening for enabling electrochemical reaction and ion drift. The reversible polarity change is robustly reproduced under 100 times of polarity switching operations, as shown in supplementary figure S2. Supplementary figure S3 also shows the evolutions of $I_d$ during the periods of polarity-switching $V_{ds}$ pulses. Gradual increase (decrease) in $I_d$ with time is observed in n-to-p (p-to-n) switching, which is consistent with the higher conductivity of p-type Te than that of n-type $MoS_2$.

In contrast to other commonly used source/drain electrode materials, Te is known to be electrochemically active. In addition, Te anions are also shown to be mobile in various solids. These two materials properties are key in enabling Te filament-based resistance switching, as previously reported[22,23]. To explore the mechanisms behind the unique mempolar phenomenon, we fabricate two control devices with Ti and Pt source/drain electrodes that are comparatively inert. Similar measurements on their transfer characteristics before and after the applications of large (+40 V, 30 s) $V_{ds}$ pulses are performed. As shown in supplementary figure 4, polarity conversion occurs in neither device.

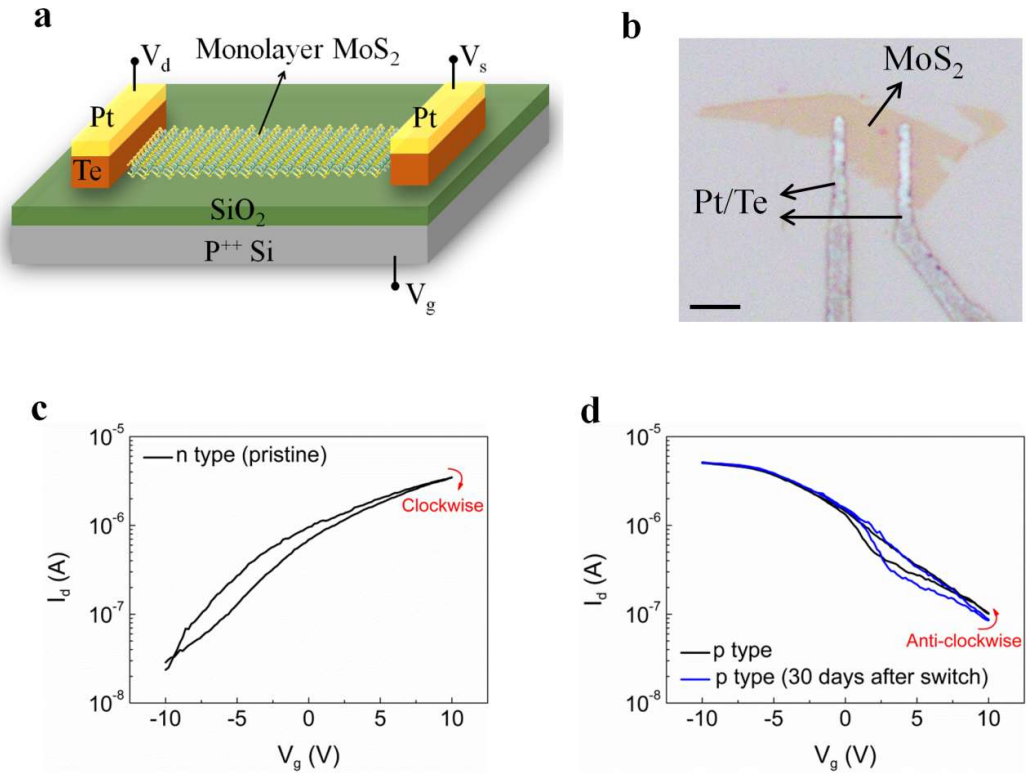

**Fig. 1 Schematic and electrical performance of the mempolar transistor. a.** Schematic and **b.** optical image of the mempolar transistor (scale bar: 2 μm). **c.** n-type transfer curve of an as-fabricated device. **d.** p-type transfer curves of the device after polarity switching.

To further identify if the atomic constitution of the channel of the mempolar transistor is changed after polarity conversion, Auger electron spectroscopy (AES) analyses of elements in the channel area for the as-fabricated n-type device, the p-type device after polarity conversion and the restored n-type device are conducted, as shown in figure 2a-c. For the as-fabricated device, it is seen that only Mo and S elements exist in the channel and no observable Te element is found. However, a remarkable amount of Te can be observed in the channel area after polarity conversion. Its distribution looks uniform along the channel width and length except, unsurprisingly, an obvious enrichment near the Te electrodes. The formation of uniform Te sheet in the channel area verifies the hypothesis that polarity conversion is due to electrochemically induced Te inclusion in $MoS_2$. When the device is switched back to n-type, the concentration of Te in the channel area dramatically decreases, indicating that Te atoms are electrochemically extracted from the channel.

Figure 2d, e show the atomic force microscopy (AFM) images of the source and drain electrodes before and after the device has undergone polarity conversion. Notches at the electrodes can be clearly seen, which can be understood as related to electrochemical reduction during n-to-p conversion and incomplete Te replenishment during p-to-n conversion. In line with the deformation of electrodes, AFM images of a local area in the channel for device before and after polarity conversion reveal that the as-exfoliated flat $MoS_2$ surface (figure 2f) turns into a pretty rough surface as the device has been converted to a p-type transistor (figure 2g). This results from Te

inclusion. Sizes and heights of the protrusions in sight can be as great as 60 μm and 6 nm, respectively. These protrusions disappear after the device has been converted back to an n-type transistor (figure 2h).

To identify the chemical structures of the channel area for the as-fabricated device and the converted p-type device, Raman spectroscopic studies are carried out. As seen in figure 2i, the as-fabricated $MoS_2$ channel shows two characteristic peaks at ~ 380 cm$^{-1}$ and ~ 405 cm$^{-1}$, in consistence with those of the in-plane Mo-S vibrational mode ($E^1_{2g}$) and the out-of-plane S-S vibrational mode ($A_{1g}$) in monolayer $MoS_2$, respectively[26]. No Te-related peak is observed. However, characteristic peaks corresponding to basal plane vibration ($A_1$) and bond-stretching vibration ($E_2$) of Te chains at 123 cm$^{-1}$ and 140 cm$^{-1}$, respectively[27], emerge after the device has been converted to a p-type transistor (figure 2j). These results further support the conjectured polarity conversion mechanism (schematic diagram in supplementary figure S5).

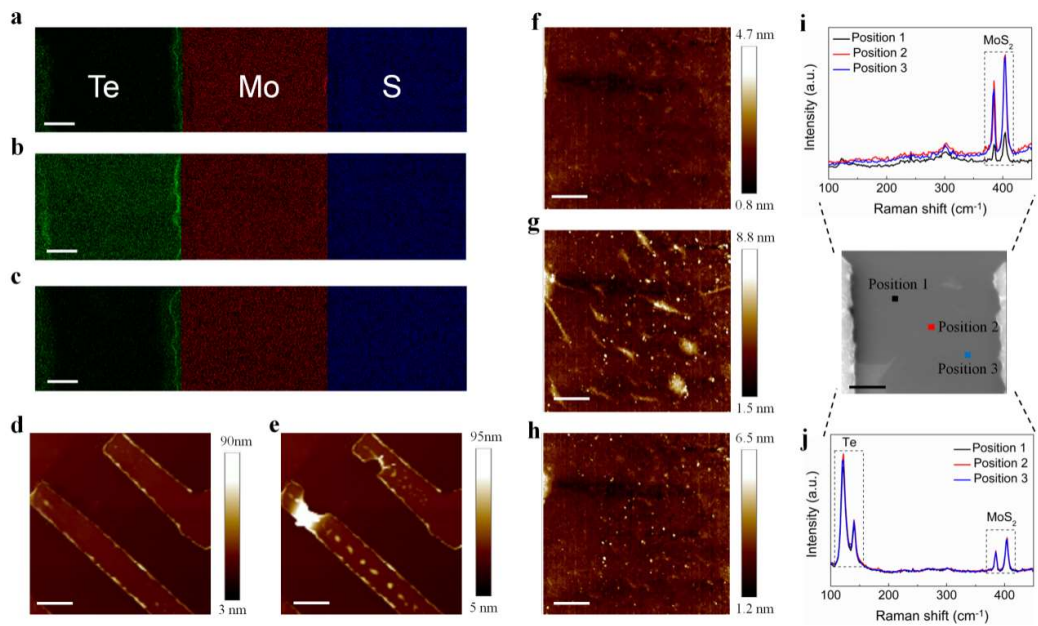

**Fig. 2 Materials characterizations of the mempolar transistors.** AES elemental mapping images of the channel area of a mempolar transistor **a.** before polarity switch, **b.** after n-to-p polarity switching and **c.** after being switched back from p-type to n-type. Scale bars for a-c: 500 nm. AFM images of the mempolar transistor **d.** before and **e.** after polarity switching (scale bar: 1.5 μm). Closed-up AFM image of the channel area of the mempolar transistor **f.** before polarity switching, **g.** after n-to-p polarity switching and **h.** after being switched back from p-type to n-type (scale bar: 120 nm). Raman spectra collected from three different areas labeled in the SEM image (scale bar: 500 nm) of the channel area of the mempolar transistor **i.** before polarity switching and **j.** after n-to-p polarity switching.

## 2.2. A TCAM cell made of two mempolar transistors

Transistors with reconfigurable polarities find applications in reconfigurable logic circuits[20], neuromorphic circuits[20,28], secure circuits[21], and so on. Here, we present a new application of mempolar transistors, that is, making TCAM cells. TCAM is a specialized type of computer memory used in certain very-high-speed searching applications. It is considered as an opposite of the more widely known random access memory (RAM). In a RAM, the user supplies a memory address and the RAM returns the data word stored at that address. By contrast, a TCAM is designed such that the user supplies a data word and the TCAM searches its entire memory based on pattern matching to see if that data word is stored anywhere in it, just like associative memory in the brain. The term "ternary" refers to the ability of the memory to store and query data using three different inputs: 0, 1 and X. The "X" input, which is often referred to as a "don't care" state, enables TCAM to perform broader searches, as opposed to binary CAM, which performs exact-match searches using only 0s and 1s.

TCAM is much faster than RAM in search-intensive applications. However, there are cost disadvantages to TCAM. Unlike a RAM that has simple storage cells, conventional TCAMs normally have more complex circuits with large physical sizes and increased power dissipation. Specifically, a single TCAM cell based on standard CMOS transistor technology requires 16 transistors. Though TCAMs based on the emerging nonvolatile memory device (or simply, memristor) technology have simpler 2T-2R cells, each employing two transistors and two memristors[29,30], the memristors must be integrated via the back-end-of-line process and thus the electrical parasitics is worsened. In this regard, our mempolar transistor as the product of the fusion of transistive and memristive functions may further simplify the structure and fabrication of the TCAM cell.

The schematic diagram of our proposed two mempolar transistor-based TCAM cell and its workings are shown in figure 3a. In this schematic, 'ML' refers to the match line which will get charged up to the supply voltage $V_{dd}$ (1 V) before the search operation. WL1 and WL2 are the two write lines through which polarity switch voltages ($\pm 30$ V) are applied to the respective mempolar transistors during the memory encoding stage. They also serve as the paths for ML discharging when a match is detected during the search stage, as will soon be introduced. For the storage of the 1 and 0 states, complementary polarity configurations are written (encoded) into the two mempolar transistors. For the storage of the X state, both mempolar transistors are written to the same polarity configuration, either p-type or n-type. SL and $\overline{\text{SL}}$ are two search lines through which the searching signal and its inverse are applied to the gate terminals of the respective mempolar transistors. The searching signal is presented as either +10 V or -10 V voltage bias, representing data 1 or 0, respectively. With the above encoding and search schemes, if a match between the searching signal and the stored memory state is detected, both mempolar transistors are in their ON states, which discharges the ML to the ground. However, if a mismatch is detected, both mempolar transistors are in their OFF states and thus the ML stays high. In the case that the TCAM cell is in the X state, no matter what the searching signal is, one mempolar transistor must be in the ON state while the other is turned off, thereby discharging the ML.

A proof-of-concept TCAM cell made with two mempolar transistors are shown in figure 3b. These two devices are fabricated from $MoS_2$ layers exfoliated on two different silicon dies which are then connected with copper wires by elargol. We experimentally demonstrate the storage of data 0/1 in this TCAM cell by converting the polarity of one mempolar transitor (mempolarT1)/(mempolarT2) to p-type while keeping the other (mempolarT2)/(mempolarT1) unchanged. Alternating 0 and 1 search data are then fed into the cell through a pair of SL and $\overline{SL}$. For each memory state, the evolution of the resistance between the ML and the ground during the search process is shown. Large resistance indicates that both mempolar transistors are in the OFF states and therefore a mismatch is detected; on the other hand, small resistance indicates that one of the two devices is turned on, corresponding to a match. The resistance contrast retains as high as $10^2$ in one thousand searches (figure 3c, d). Unlike the recently reported ferroelectric TCAM cell[31] in which both writing and searching signal transductions share the same pathway (i.e., SL/$\overline{SL}$), our cell uses two signal transduction pathways for these two operations, i.e., the SL/$\overline{SL}$ for searching and the WL1/WL2 for writing. The decoupling of the search and write operations prevents the stored state from being disturbed by the searching signals and therefore may enable long retention time and enhance reliability.

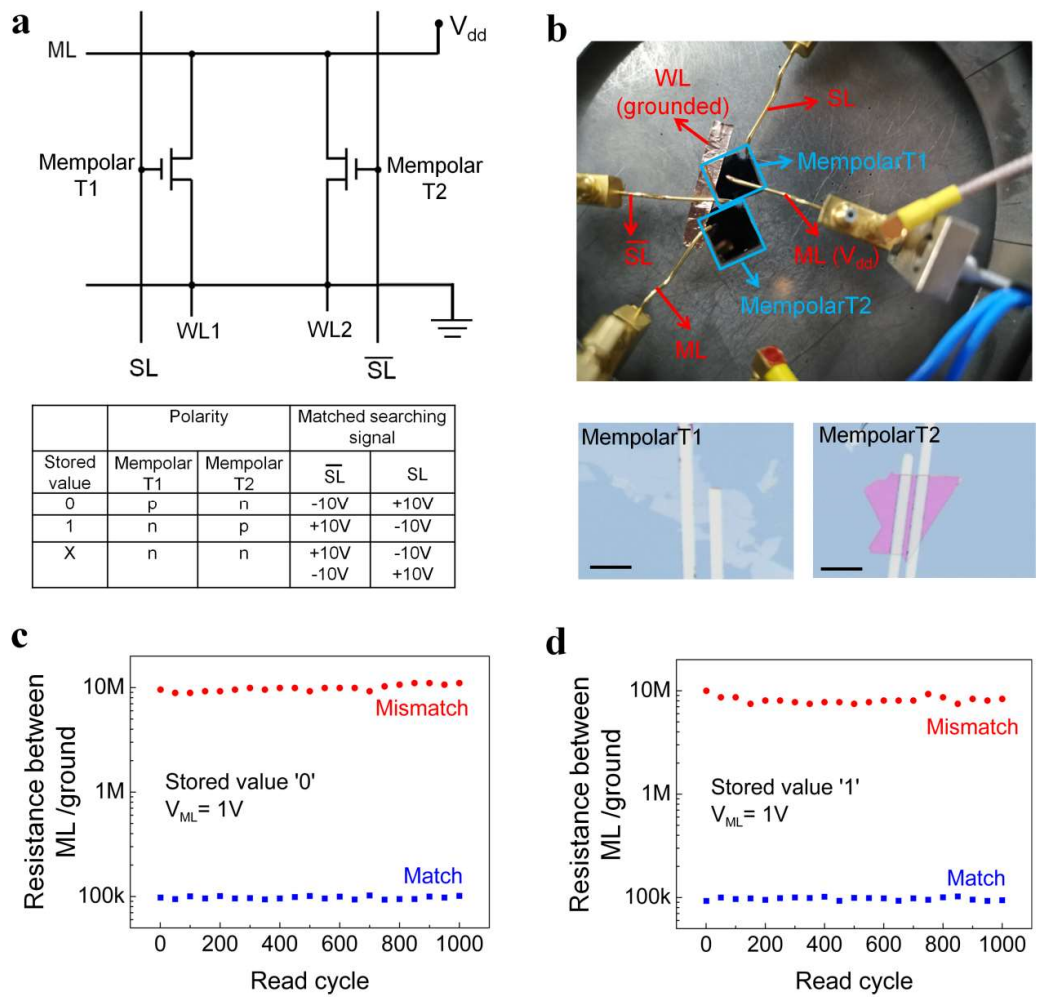

**Fig. 3 Two-mempolar transistors-based TCAM cell. a.** Circuit diagram of the two-mempolar transistors-based TCAM cell during the search stage and its operating mode. **b.** A proof-of-concept two-mempolar transistors-based TCAM cell (scale bar: 2 μm). Evolutions of the measured resistance between the ML and the ground with the searching signals alternating between matched and mismatched signals when **c.** '0' and **d.** '1' are stored in the TCAM cell.

## 2.3. Mempolar transistor-inspired method for regularizing neural networks

As mentioned before, the mempolar transistor with n (p) polarity can show certain degree of clockwise (counter-clockwise) hysteresis in its transfer curve (figure 1c and 1d), which can be attributed to trapping/detrapping of carriers in/from the gate oxide during the course of $V_g$ sweeping. As seen from the transfer curves in figure 4a (figure 4b), the channel conductance (under $V_g=0$ V) of the mempolar transistor with n (p) polarity keeps decreasing (increasing) with successive forward and backward voltage sweepings between 0 V to +10 V. Pulse measurements are also carried out, revealing similar trends of conductance changes in devices with n and p polarities, respectively (supplementary figure S6). These phenomena have been widely exploited by neuromorphic engineers for emulating the long-term plasticity of synapses in the training phase of neural networks[16].

As shown in figure 4c, 10 cycles of forward and backward $V_g$ sweeps between 0 V to +10 V applied to an as-fabricated n-type device (1→2) lead to dramatic decrease in its baseline $I_d$ ($I_d$ under $V_g=0$) from about 0.7 μA to 100 nA. After the induction of this long-term synaptic depression (LTD), a positive polarity switching $V_{ds}$ is applied (2→3). The resulting p-type device has large baseline $I_d$ about 10 μA. Recall that polarity conversion is elicited by applying a voltage across the source and drain terminals. Therefore, this operation, in principle, should not influence the charge trapping state of the gate oxide which only depends on the history of gate inputs. To verify that the charge trapping state of the gate oxide is not influenced by polarity conversion, we apply a negative polarity switching $V_{ds}$ to convert the device back to n-type (3→4). It is seen that the transfer curve of the present n-type device overlap that of the long-term depressed device before polarity conversion. Likewise, long-term synaptic potentiation (LTP) is induced in the p-type device converted from the as-fabricated n-type device, followed by p-to-n conversion and then n-to-p conversion, as shown in figure 4d. Results from these measurements consistently indicate that polarity conversion does not influence the charge trapping state of the gate oxide but can give rise to reverse effect to what long-term plasticity induces before polarity conversion. Specifically, the strength of the reverse effect is monotonically dependent on the pre-accumulated long-term plasticity.

Drawing inspiration from this device behavior, we propose a new algorithm 'FlipWeight' and demonstrate its application by simulations in the context of regularizing neural networks. Traditionally, training a deep neural network (DNN) that can generalize well to new data is a challenging task. This is because a typical DNN has so many parameters (over-parameterized) and limited available data, exhibiting a significant tendency toward overfitting on the training dataset. Approaches to reduce error in generalizing to out-of-sample data points are referred to as regularization methods[32]. Generally speaking, the rationale behind regularization is constraining the complexity of the DNN model by either reducing the number of synaptic connections or reducing values of synaptic weights. Dropout[33,34] and its variant DropConnect[35] are two of the most widely used regularization methods. As their names suggest, these two methods randomly drop a number of neurons or connections for each batch during training. With this, not all but only a fraction of weights are updated. After the training on a batch of samples, the omitted neurons (and connections attached to it) or omitted connections in the last batch of training are recovered. A new set of neurons or connections are randomly selected and omitted in the next training batch. Unlike these two methods, our method adapts the idea of reducing values of synaptic weights to improve generalization performance.

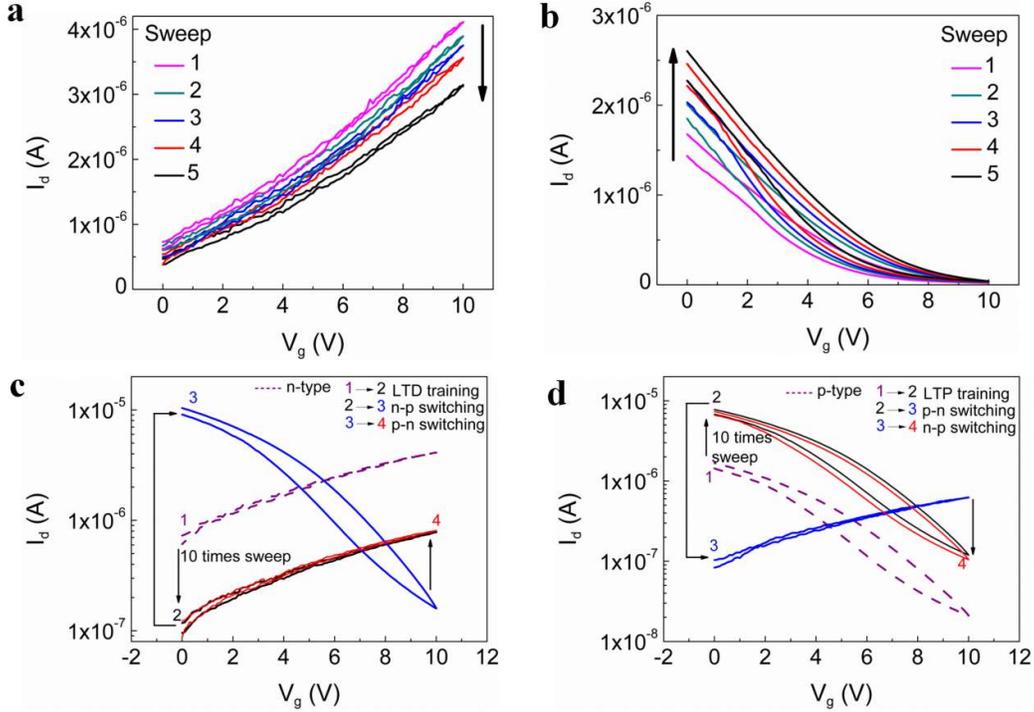

**Fig. 4 Gradual channel conductance changes in mempolar transistors and flipping between the high and low conductance states via polarity switching. a.** Gradual decrease in the baseline $I_d$ ($I_d$ under $V_g=0$) of an n-type mempolar transistor under successive positive $V_g$ sweep. **b.** Gradual increase in the baseline $I_d$ of a p-type mempolar transistor under successive positive $V_g$ sweep. Flipping back and forth between the high and low conductance states **c.** in an n-type mempolar transistor before which the channel conductance has been gradually tuned to be relatively low and **d.** in a p-type mempolar transistor before which the channel conductance has been gradually tuned to be relatively high.

Large weights in a DNN are a sign of a complex network that has a tendency to overfit the training data[36]. Therefore, we consider randomly selecting large weights that are over a threshold and scaling them down according to a certain rule before each training batch starts. A weight scaling rule that naturally matches the properties of our mempolar transistor is inverse scaling with respect to the initial weight value. After this pre-treatment, the standard backpropagation (BP) method is used to calculate the gradients for all the weights in the DNN. Except those pre-treated weights, all the other weights are directly updated using the calculated gradients. The pre-treated weights are first recovered to their original values (also supported by device functions) and then updated using the corresponding gradients calculated from the pre-treated DNN. In the next training batch, a new set of large weights are randomly selected and scaled down, followed by the same procedure of BP calculations and weight update.

We benchmark this FlipWeight method against the Dropout and DropConnect methods on a five-layer convolutional neural network (CNN), as shown in figure 5a. We point out that the FlipWeight method is only used in the fully-connected (FC) layers, like Dropout and DropConnect. As presented in figure 5b and 5c, CNN trained with either Dropout or DropConnect or FlipWeight technique can achieve higher validation accuracies and lower loss values compared to the baseline that is implemented without any regularization technique. The gaps between the training curves and the validation curves are also reduced dramatically with the use of these regularization techniques, verifying their effectiveness in improving the generalization performance. Notably, although the final validation accuracy and loss value of CNN implemented with FlipWeight method are similar to those of the CNNs implemented with Dropout and DropConnect, it is evident that our FlipWeight method results in faster convergence than do the other two methods. This can be understood from the fact that all the weights are updated in each training iteration in CNN regularized by our FlipWeight method, while connections omitted in the Dropout or DropConnect approach are simply not involved in weight updating.

We also compare the mean, standard deviation and sum of absolute value of weights between CNN models regularized by different methods, as shown in figure 5d-f. It is seen that our FlipWeight method leads to overall excitatory connections (positive mean) while the connections in baseline CNN without regularization and CNNs regularized by Dropout and DropConnect are overall inhibitory (negative mean). Our FlipWeight method also leads to the largest standard deviation of weights and sum of absolute values of weights. The weight distributions are also visualized. The truncated Gaussian distribution for weight initialization is shown in figure 5g. After training, the baseline model without regularization shows a wide distribution of weights among which many have large values (figure 5g). For Dropout and DropConnect, the weight distributions are tighter and peak near zero (figure 5h, i). In stark contrast, the weight distribution in CNN regularized by FlipWeight method is multimodal and even tighter, whose several peaks are close to zero (figure 5j). These results demonstrate the uniqueness of our FlipWeight method that can achieve state-of-the-art performance.

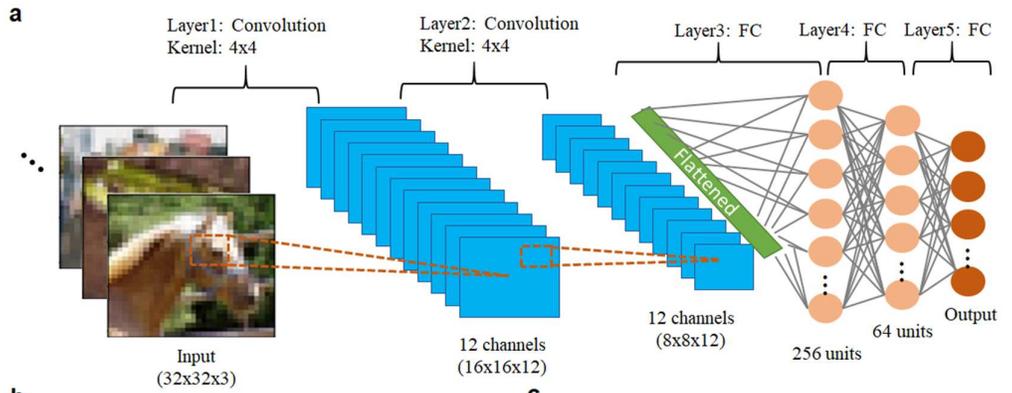

**a**

Layer1: Convolution
Kernel: 4x4

Layer2: Convolution
Kernel: 4x4

Layer3: FC   Layer4: FC   Layer5: FC

Flattened

Input
(32x32x3)

12 channels
(16x16x12)

12 channels
(8x8x12)

256 units

64 units

Output

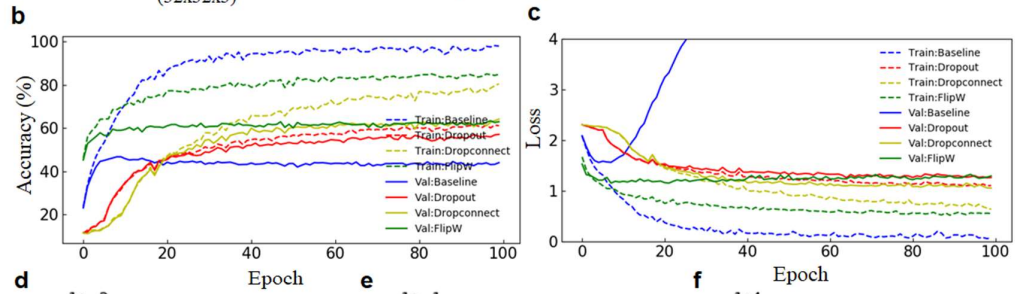

**b**

**c**

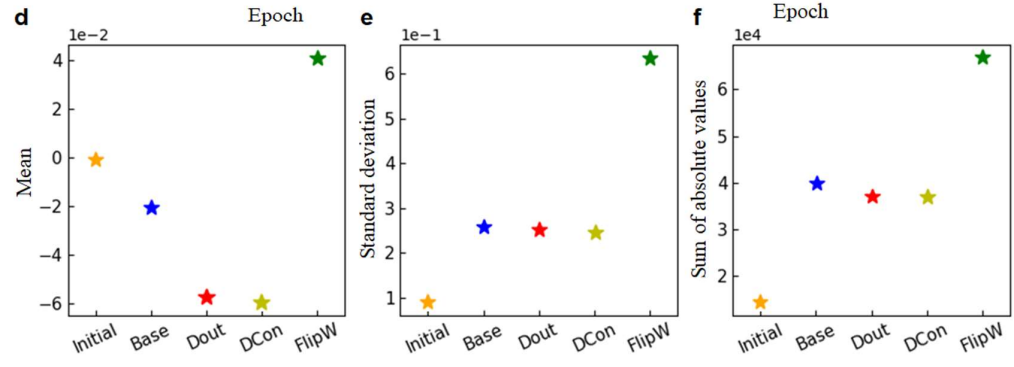

**d**

**e**

**f**

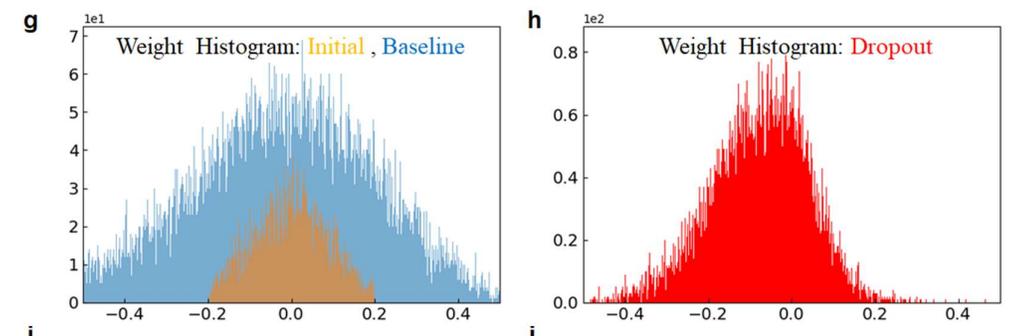

**g**

**h**

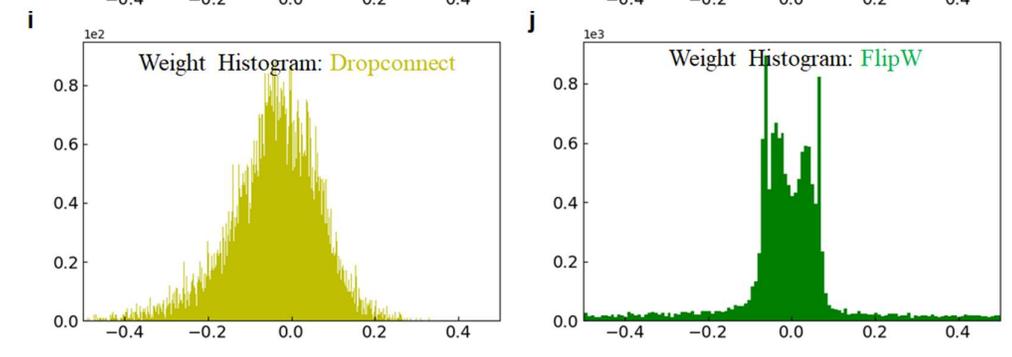

**i**

**j**

**Fig. 5 Performance and characteristics of the CNNs regularized by different methods. a.** Schematic illustration of the adopted network structure for image recognition. Comparison of the **b.** convergence curves and **c.** loss curves obtained from unregularized model and models regularized by different methods. Comparison of the **d.** mean, **e.** stand deviation and **f.** sums of absolute values of synaptic weights among different models after training, where "Initial" denotes the untrained model, "Base" denotes unregularized baseline model, "Dout", "DCon" and "FlipW" denote models regularized by Dropout, DropConnect and FlipW methods, respectively. Comparison of weight distribution among **g.** the untrained model and the unregularized baseline model, and models regularized by **h.** Dropout, **i.** DropConnect, and **j.** FlipWeight methods, respectively. 图 j FlipWeight

## 3. Conclusion

In summary, we introduce the emerging ion-based memristive functions into the purely electronic transistors and demonstrate a new type of transistor named 'mempolar transistor' whose polarity can be run-time switched between n-type and p-type in a non-volatile manner. This novel transistor function is achieved by the use of the emerging semiconducting Te as the electrochemically active source/drain contact material, in combination with monolayer $MoS_2$ channel, which results in a gated lateral $Te/MoS_2/Te$ memristor, or from a different perspective, a transistor whose channel can be converted reversibly between n-type $MoS_2$ and p-type Te. Mempolar transistors address a key drawback of the previously showcased transistors with reconfigurable polarities, that is, polarity reconfiguration was volatile and realized via electrostatic control from additional gate terminals. When used in reconfigurable logic circuits or secure circuits, mempolar transistors will potentially mitigate the problems of excessive energy consumption, extensive hardware overhead and massive interconnections. In addition to these manifest advantage, we design and demonstrate experimentally a TCAM made of only two mempolar transistors, which used to require a dozen normal transistors. We also develop and demonstrate by simulations a device-inspired regularization method for training ANNs, which achieves state-of-the-art performance. This work broadens the functionality of transistors and provides the implication that rich technological opportunities are available for the fusion between electronics and ionics.

## 4. Methods

*Device fabrication:* The $MoS_2$ flakes were mechanically exfoliated from bulk crystals (purchased from Six Carbon Technology, Inc.) onto the $SiO_2$ (300 nm)/Si substrate. The source and drain electrodes made from 50 nm Te and 20 nm Pt protective layers were deposited by magnetron sputtering after the standard electron-beam lithography patterning.

*Electrical measurements:* Cyclic quasi-DC voltage sweep measurements were performed by the Keysight B1500A semiconductor analysis system. The Keysight B1530A waveform generator/fast measurement unit is used to perform the pulse measurements.

*Materials characterizations:* The AES analyses were performed by a scanning auger microprobe (PHI710, ULVAC). The morphology characterizations of the devices were performed by an AFM (DIMENSION ICON, BRUKER) in ScanAsyst mode. The Raman spectra were obtained on a single-gating micro-Raman spectrometer (Horiba-JY T64000) excited with 532 nm laser.

*Neural network model parameterization, training and tests:* We evaluated the performance of various regularization methods in CIFAR-10 pattern classification tasks[37]. The adopted network structure was [Input-12C4-12C4-768FC-256FC-64FC-10] (C: convolution, FC: fully-connected layer). We modelled the flipping of a connection weight from its current large value to a small value according to the physical process of polarity switching-induced channel conductance change in the mempolar transistor and by simplifying this process as obtaining the reciprocal of the initial large value (inverse scaling of weight). All CNN models were trained using the adaptive moment estimation (Adam) optimizer[38] for 160 epochs with batch size of 100 and an initial learning rate of 0.0005. The retaining proportions for models regularized by Dropout or DropConnect were set to 0.5. For the FlipWeight method, connections in the FC layers with weights larger than a threshold value are inversely scaled to their reciprocals with the effect of a scaling factor before each training epoch starts. 0.06 was found to be a suitable threshold value. The hyper-parameters were kept the same for all tested models in this work. The simulations were performed by Tensorflow1.15.0 on 4 RTX 2080Ti GPUs.

## Supporting Information
Supporting Information is available from the author.

## Acknowledgments
Y. Y., L. X. and M. X. contributed equally to this work. H. L. conceived the idea. Y.Y. and L.X. performed the device fabrication and measurements under the supervision of H.L. and W.D.. M.X. conducted the neural network simulations under the supervision of J.P.. H. Liu and D.L. assisted the device fabrication. Y.Y., M.X. and H.L. wrote this manuscript. This research was supported by National Natural Science Foundation (grant nos. 61974082, 61704096, 61836004), National Key R&D Program of China (2021ZD0200300, 2018YFE0200200), Youth Elite Scientist Sponsorship (YESS) Program of China Association for Science and Technology (CAST) (no. 2019QNRC001), Tsinghua-IDG/McGovern Brain-X program, Beijing science and technology program (grant nos. Z18110001518006 and Z191100007519009), Suzhou-Tsinghua innovation leading program 2016SZ0102, CETC Haikang Group-Brain Inspired Computing Joint Research Center.

## Competing interests
The authors declare no competing interests.